# On M2M communications standards for smart metering

Manfred Sneps-Sneppe[1], Anatoly Maximenko[2] ,Dmitry Namiot[3]

[1] Ventspils University College
Ventspils International Radioastronomy Centre
Ventspils, Latvia,
{ manfreds.sneps@gmail.com }

[2] National Research University of Electronic Technology
Zelenograd Center for Innovation and Technology
Moscow, Russia
{ maximinus@bk.ru }

[3] Lomonosov Moscow State University
Faculty of Computational Mathematics and Cybernetics
Moscow, Russia
{ dnamiot@gmail.com }

**Abstract.** The paper discusses M2M communications standards for smart metering. One of the our goals is to show the failures of ETSI standartization process for M2M communications. Our paper proposes some extesions to ETSI standards. At the first hand, it is M-Bus protocol and Open Metering System based on M-Bus. The paper shows how to estimate wireless M-bus throughput and how to avoid collisions. After analysis of Open API for M2M, submitted to ETSI, we propose a new approach in the client-side web development – Web Intents. The main goal for our suggestions is to simplify the development phase for new applications by support asynchronous calls and JSON versus XML for data exchange.

## 1. Introduction

The M2M (Machine-to-Machine) communications industry has attracted increasing attention during the past several years. Analysys Mason forecasts [1] that the number of M2M device worldwide connections will grow from 62 million in 2010 to 2.1 billion devices in 2020, at a 36% year-on-year growth rate. Household use (utilities, power meters, etc.) constitutes an over helming fraction – 62%. According to the prognosis [1], over 90 percent of forecast connections will utilize wireless. While many M2M deployments will make use of short-range or proprietary radio links, cellular-based M2M solutions will be preferred where mobility is required, or where high data volumes or data transfer rates are involved. In order for the billions of devices, services and networks to realize the full M2M potential, standardization will be a key.

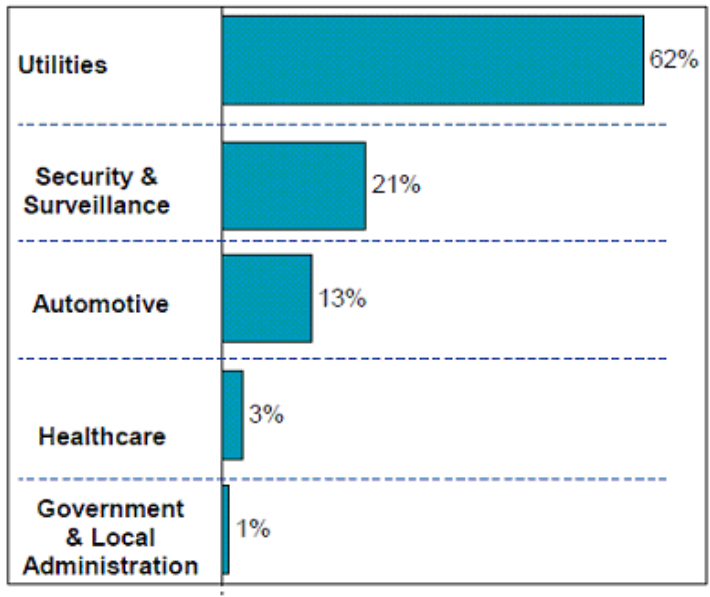

Figure 1. Commercial and consumer M2M device connections by industry sector, worldwide, 2020 (Source: Analysys Mason, 2011 [1]).

M2M architecture interfaces. Open standards, open protocols, open architecture and open web are some of the key concepts in the home automation industry, but it is rather far from real openness.

1) Many people believe that open protocol implies that if a controller fails from one vendor, they can replace it with another vendor. It is not as simple as plug and play. There are overhead vendor dependent software tools that must be used to configure and program the controller before connecting to the network.

2) The open protocol standards (e.g., the most popular standards BACnet or LonWorks) do not define a standard programming language or rules to program an application controller. The actual interpretation of the code written by the user is proprietary to each vendor.

3) In the most cases, the manufacturer and/or the system integrator will not allow the facility manager to view the programs.

Several standartization bodies are active in the field of M2M communications (e.g. IETF, ETSI, OMA, BBF, OSGi, HGI, etc.) [2, 3]. In the paper, we follow the EURESCOM concept of the M2M architecture [4 ]. There are seven basic interfaces of M2M architecture (Figure 2).

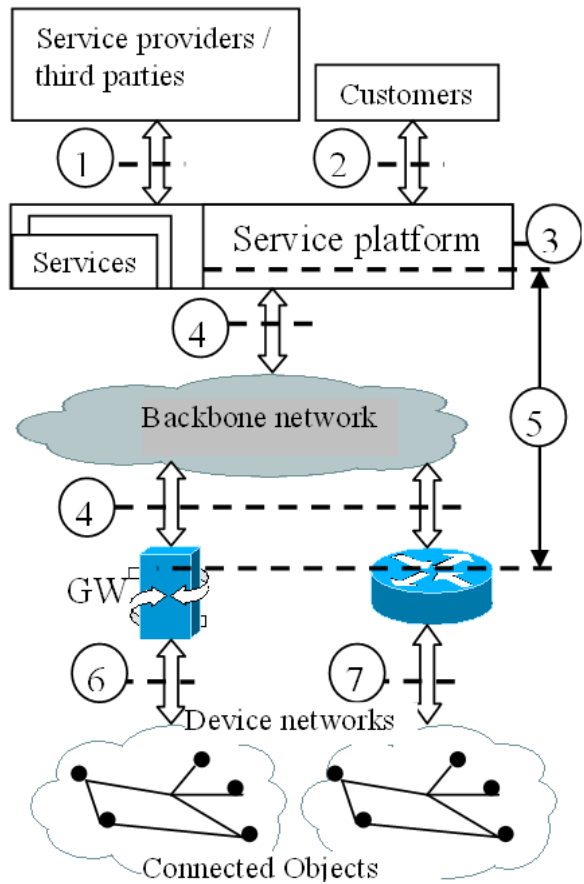

Figure 2. An M2M architecture instance [4].

Interface 1 - between the platform and external service providers running their services remotely;

Interface 2 - between the platform and the customer applying the features offered by the platform, may be accessed conveniently via a standard web browser;

Interface 3 - for installation support, access to remote databases, remote management, etc.;

Interface 4 - to the backbone IP network;

Interface 5 is the application level interface between the service platform and Connected Objects on the device side. (This is described in more detail in the OpenAPI section below.)

Interface 6 is application specific (non-IP based, e.g., M-bus, ZigBee, LON, Zwave) and requires gateway.

Interface 7 may be identical to Interface 4. However, it may be optimized for M2M applications according to specifications from the IETF (i.e. from the working groups 6lowpan and Roll).

All software interfaces 1 to 7 must be open, as requested by EC Mandate M/441, but up to now ETSI efforts are restricted by Open M2M API only (i.e., Interface 1). The Open API for M2M applications developed in EURESCOM study P1957 [4] have been submitted as a contribution to ETSI TC M2M [5] for standardization.

Our main goal is here twofold. Besides the describing the current state of M2M communications standards for smart metering, we try, at the first hand, to estimate the place of our proposal to promote M-Bus protocol [6, 7] as a unique protocol between tens of existing ones. The second point relates to some new additions in M2M APIs architecture [8, 9, 10]. Namely, we are going to propose Web Intents as add-on for the more traditional REST approach in order to simplify the development phases for M2M applications. The key moments in our proposals are: JSON versus XML, asynchronous communications and integrated calls.

The rest of the paper is organized as follows. In Sections II and III we refer to ETSI standardization mandate M/441 for smart metering and ETSI's 3rd workshop on M2M communications. Section IV contains the description of Open Metering System specification developed by German companies ZVEI, FIGAWA and KNX. Sections V and VI consider Open API for M2M, submitted to ETSI and a never web tool – Web Intents for enhancement of M2M middleware.

## 2. Smart metering standardization mandate M/411

Considering M2M communications as a central point of Future Internet, European commission creates standardization mandate M/441 [11]. The mandate M/441, issued on 12th March 2009 to CEN, CENELEC and ETSI, is a major development in shaping the future European standards for smart metering and Advanced Metering Infrastructures. The general objective of the mandate is to ensure European standards that will enable interoperability of utility meters (water, gas, electricity, heat).

As it was pointed out in the MOU of Mandate, there are about 110 applicable technical standards available today which cover parts of a Smart Metering application. No standard covers the full application range. Many component-level standards for M2M communications already exist, addressing various radio interfaces, different meshed or routed networking choices, or offering a choice of identity schemes. Each is optimised for a particular application scenario and there is therefore a degree of fragmentation. Until now, little effort has been made to bring all these pieces together, and identify the standardization gaps which exist. ETSI is now confronting to that unsolved problem.

According to Mandate, standardization bodies are request to develop European Standards for:
1. A software and hardware open architecture for utility meters – the open architecture must support secure bidirectional upstream and downstream communications through standardized interfaces, scale to support simple and the most complex application, consider current and future communication media, and management and control systems for consumers and services providers;

2. Solutions for additional functionalities within the interoperable framework – additional functionalities within the interoperable framework must be achieved to enable full interoperability solutions;

3. The standards to be developed must permit innovation in the protocols that enable remote reading of utility meters, advanced information and management services and fully integrated instruments, modular and multi-part solutions.

In order to achieve full interoperability, as requested by Mandate M/441 [11], and with the OSI model as a reference, open interface standards must be defined for all layers of the communications protocol stack that reside on the meter, both upstream and downstream (e.g. for all interfaces as Figure 2 shows).

## 3. On M2M standartization: a failure case

Mandate M/441 has gone in power three and a half years ago, and what have we now? ETSI's 3rd workshop on Machine to Machine (M2M) communications, held in Mandelieu, France on October 23-25, 2012, gathered leading experts from all over the world to hear how ETSI M2M technology standards are being deployed. "With 270 registered delegates from four continents, 25 speakers, thirteen live demonstrations of M2M-based applications and two days of intense discussion, this year's event was again a success," – the official workshops' site assure us [12]. In reality, the state-of-the-art with M2M standards is far from hopeful.

The demos of M2M-based applications involving 28 companies were carried on.The demos covered once again a respectable cross section of the application domains such as: Smart Metering, Home automation, Energy Efficiency, Smart building, Smart City, Smart Parking, Exercise, Gaming and Home Energy, Management Systems linked with Social Networking Service and others. But the existing standards IETF CoAP and 6lowpan, ETSI M2M, OMA DM, BBF TR069, OSGi, HGI, etc. and the existing protocols ZigBee, KNX, etc. are far from convergence.

Now the new international M2M Partnership Project "oneM2M" has started. The list of funding partners include ETSI (Europe), ATIS and TIA (US), CCSA (China), TTA (Korea), ARIB and TTC (Japan). But, we guess, the leading role of ETSI stays no easier.

The title of paper "M2M: An Ecosystem in Flux" is rather symptomatic [13]. As before (in the power of M/441) ETSI TC M2M wants to provide an end-to-end view of M2M standardization. Many disjointed standards have already been developed, but no one has successfully brought all these pieces together.

TIA has recently emerged as a leading voice in the U.S. and could be the best hope in the framework of the „oneM2M" project. Regardless of the geography of the „oneM2M" network or the device, the continued proliferation of connected devices on a global scale will force everyone to agree on one standard sooner rather than later.

The China's assistance is one more hopeful sign. "It is urgent to develop national-level standards, even for the three major Chinese mobile operators that are developing different M2M standards," the representative of the Science and Technology Committee of China Telecom says. "Only when standardization is achieved can M2M modules and terminals fulfill large-scale and cost-effective manufacturing, and the industry value chain match the potential of M2M services."

## 4. Open Metering System: a challenge

The 3d ETSI M2M workshop demos were represented by many companies (e.g. smart metering from Gridpocket, Sierra Wireless and LAAS-CMRS, M2M gateway from Actility and NEC, etc.). But these demos were about interoperability only and not much about convergence of standards. Meanwhile, leading meter manufacturers and technology providers in Europe have joined the effort to create the new open standard for metering. The new Open Metering System (OMS) specification has been developed to meet a demand for interoperable solutions for meter reading, and a unified approach for the different media (electricity, gas, heat and water). In 2009, the three-part specification was released [14]. The Open Metering System alliance was initiated by German companies ZVEI, FIGAWA and KNX.

The new specification is based on established norms and standards where it has been possible. The tertiary communication is solely based on TCP/IP, and the primary communication is based on the M-Bus standard (wired or wireless), EN 13757.

For wired connections the physical layer M-Bus according to the European standard EN 13757 is proposed. It is a two-wire system, optionally capable of providing power to multi-utility (battery-powered) meters (Figure 3).

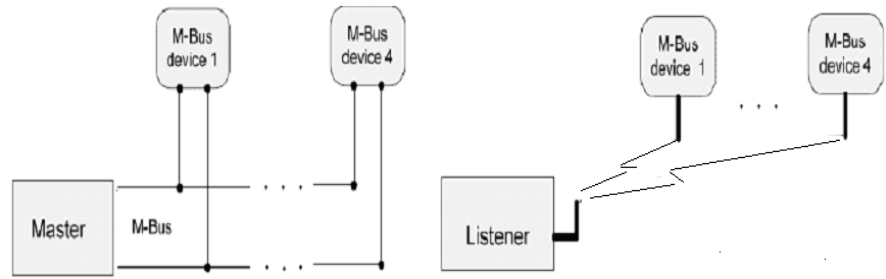

Figure 3. M-bus wired and wireless infrastructure with 4 meters.

The specification defines a Multi Utility Communication (MUC) device, which acts like an intelligent data concentrator between the automated meter management (AMM) back office system (for billing or other purposes), and the metering and actuator devices. The MUC can be integrated into a meter (typically an electricity meter) or it can be a standalone unit. Figure 4 shows a simplified Open Metering System overview.

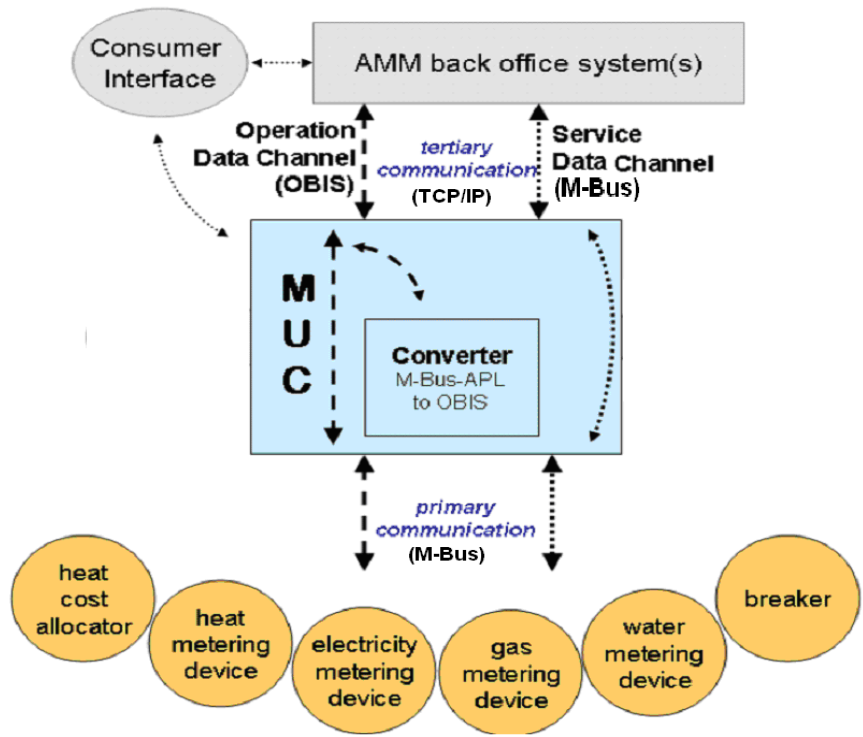

Figure 4. Open Metering System overview

The primary communication is between meters and the MUC. A lot of effort has been put into unifying this part to support all media, as well as actuators and displays. The secondary communication is defined as an extension of the primary communication using simple repeaters or a multi-hop routing protocol. The tertiary communication is between the MUC and the back office AMM system.

The specified data format is OBIS (Object Identification System) coded values [15]. The wired/Wireless M-Bus link to the meter supports both OBIS (not shown), as well as the M-Bus application data format (EN 13757-3). The MUC will translate the M-Bus application data format into OBIS before it is sent to the AMM on the operation data channel. A service data channel from the MUC to the AMM supports M-Bus formatted data as well.

Several companies (e.g., Radiocrafts, Sierra Wireless, Telit, Amber-wireless, etc.) introduced Open Metering System compliant gateways for Smart Meters to market. But they did not demonstrate them at ETSI M2M workshop.

## 5. On Wireless M-Bus throughput: future work

Of course, there are some OMS problems not solved yet. One hard question is - how to estimate wireless M-bus throughput and how to avoid collisions.

Wireless M-Bus has been optimised for low battery consumption and short duty cycles. It provides various modes to support different scenarios like stationary and mobile readout. The main direction of the traffic is from the meter to the data collector system, with a back channel for commands, especially when the meter is battery powered. In the case of mains powered meters, there are fewer limitations.

Standard EN 13757 describes several variants for wireless meter communications (S1, T1, S2, T2). They cover all types of meter communication including mobile and stationary readout modes. A smart meter scenario requires a stationary receiver and frequent transmission of meter data to support user consumption feedback and variable tariffs.

S1 and T1 are unidirectional standards where the meter frequently (seconds to hours) transmits telegrams containing meter identification together with metered data. S2 and T2 are compatible bidirectional enhancements of S1 respectively T1. Both enable a communication hub to multi-utility meter communication after each multi-utility meter to communication hub telegram. Figure 5 shows the intervals of transmission for different applications and metering media [16].

| Meter Type | Mandatory billing and visualisation | | Informative aspects (consumer) |
|---|---|---|---|
| | Average Send interval Max (min) | Visualisation interval for energy provider (h) | Visualisation interval for consumer (min) |
| Electricity | 7,5 | 1 | 15 |
| Gas | 30,0 | 1 | 60 |
| District Heating | 30 | 1 | 60 |
| Hot Water | 240 | 24 | - |
| Repeater | 240 | - | - |

Figure 5. of transmission for different applications and metering media.

Let us give an insight in wireless metering technique. Aloha, the simplest form of random uncoordinated channel access mechanism, allows any packet that is available for transmission to be sent on the medium with successful packet reception reported via an acknowledgement. In the absence of an acknowledgement, the transmission is repeated.

In pure Aloha, the probability of successful packet transmission of fixed length packets of duration m is the probability that no other packet is transmitted in the interval 2m. In other words, in its simplest form, Aloha assumes all non-collided packets are successfully delivered.

The number of transmission attempts obviously depends on the number of active nodes and the probability with which each node generates a packet. In the classical Aloha analysis, inter-arrival is a Poisson distribution as each transmission is assumed to be uncorrelated with any other. Figure 6 shows that the pure Aloha maximum efficiency is a less more than 18 %.

The "Listen Before Talk" approach enhances throughput. Clear Channel Assessment (CCA) is used to avoid collisions by sensing the medium before transmission. If the medium is occupied the transmitter defers with some randomised delay before re-attempting its transmission operation by once more sensing the medium. CCA sensing mechanism is used in Carrier Sense Multiple Access (CSMA) algorithms.

For an equivalent payload and data rate, the throughput of a CSMA Collision Avoidance (CA) mechanism is much greater than Aloha (Figure 6). If we examine the performance curves for Pure and Slotted Aloha it can be seen that they begin to diverge significantly only at offered loads of approximately 10 %.

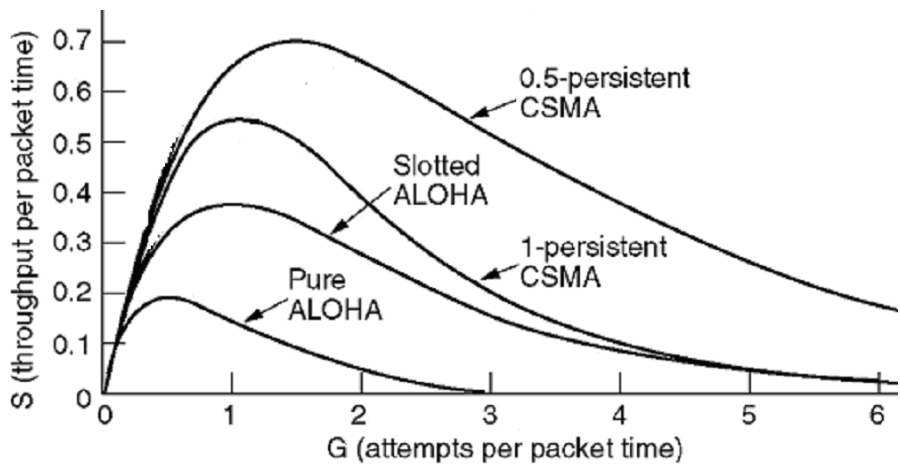

Figure 6. Comparison of Aloha and Carrier Sense mechanism.

The throughput analysis for OMS wireless mode is an issue for future work [17].

## 6. Open API for M2M

This section describes an Open API for M2M, submitted to ETSI. It is probably the most valuable achievement of Mandate M/441 at this moment. The OpenAPI for M2M applications developed jointly in Eurescom study P1957 [4] and the EU FP7 SENSEI project [18] makes. The OpenAPI has been submitted as a contribution to ETSI TC M2M [5] for standardization.

Actually, in this Open API, we can see the big influence of Parlay specification. Parlay Group leads the standard, so called Parlay/OSA API, to open up the networks by defining, establishing, and supporting a common industry-standard APIs. Parlay Group also specifies the Parlay Web services API, also known as Parlay X API, which is much simpler than Parlay/OSA API to enable IT developers to use it without network expertise [19].

The goals of Open API for M2M are obvious, and they are probably the same as for any unified API. One of the main challenges in order to support easy development of M2M services and applications will be to make M2M network protocols "transparent" to applications. Providing standard interfaces to service and application providers in a network independent way will allow service portability [20].

At the same time, an application could provide services via different M2M networks using different technologies as long as the same API is supported and used. By this way, an API shields applications from the underlying technologies and reduces efforts involved in service development. Services may be replicated and ported between different execution environments and hardware platforms [21].

This approach also lets services and technology platforms to evolve independently. A standard open M2M API with network support will ensure service interoperability and allow ubiquitous end-to-end service provisioning.

The Open API provides service capabilities that are to be shared by different applications. Service Capabilities may be M2M specific or generic, i.e., providing support to more than one M2M application.

Key points for Open API:
- supports interoperability across heterogeneous transports
- describes high-level flow and does not dictate implementation technology
- is message-based solution
- combines P2P with client-server model
-  supports routing via intermediaries

At this moment all the points are probably not discussable except the message-based decision. Nowadays, publish-subscribe method is definitely not among the favorites approaches in the web development, especially for heavy-loading projects. You can see the more detailed description in our paper [22]. Let us just highlight some discussable areas.

1.    Roadmap for the ongoing FP7 project "FI-WARE" [23] highlights the plans for M2M General Enabler (Gateway). In the same time M2M Open API can benefits from more elaborated FI-WARE persistence and cloud offerings. Note that because we are talking about server-side solution, there is no problem with so called sandbox restrictions. But it means, of course, that such kind of request could not be provided right from the client side as many modern web applications do. In the same time FI-WARE authors, at least according to roadmap, pays attention to JSON and other client-side technologies. The reasons are obvious – the modern web-based technologies can save the development time.

2.    Another area where M2M API in the current form is weak belongs to data persistence. We should keep in mind that we are talking about the particular domain – M2M. In the most cases, all business applications here will deal with some metering data. As soon as we admit, that we are dealing with the measurements in the various forms we should make, as seems to us a natural conclusion – we need to save the data somewhere. It is the core business for M2M – save data for the future processing. So, the question is very easy – can we talk about M2M applications without talking about data persistence?

As seems to us, even right now, before the putting some unified API in place, the term M2M almost always coexists with the term "cloud". And as we can see, almost always has been accompanied by the terms like automatic database logging, backup capabilities, etc.

So, maybe this question is more for the discussions or it even could be provocative in the some forms, but it is: why there is no reference API for persistence layer in the unified M2M API? Is it possible in general to create data gathering API without even mentioning data persistence?

## 7. Web Intents vs. Open API from ETSI

This section devoted to the relatively new approach in the client-side web development – Web Intents. The first time Web Intents usage for M2M applications was proposed by Dmitry Namiot [21].

Let us start from the basic. Users use many different services on the web to handle their day to day tasks, developers use different services for various tasks. In other words, our environment consists of connected applications. And of course, all they expect their applications to be connected and to work together seamlessly.

It is almost impossible for developers to anticipate every new service and to integrate with every existing external service that their users prefer, and thus, they must choose to integrate with a few select APIs at great expense to the developer.

As per telecom experience, we can mention here the various attempts for unified API that started, probably, with Parlay. Despite a lot of efforts, Parlay API's actually increase the time for development. It is, by our opinion, the main reason for the Parlay's failure.

Web Intents solves this problem. Web Intents is a framework for client-side service discovery and inter-application communication. Services register their intention to be able to handle an action on the user's behalf. Applications request to start an action of a certain verb (for example: share, edit, view, pick, etc.) and the system will find the appropriate services for the user to use based on the user's preference. It is the basic [23].

Going to M2M applications it means that our potential devices will be able to present more integrated for the measurement visualization for example. The final goal of any M2M based application is to get (collect) measurements and perform some calculations (make some decisions) on the collected dataset. We can go either via low level API's or use (at least for majority of use cases) some integrated solutions. The advantages are obvious. We can seriously decrease the time for development.

Web Intents puts the user in control of service integrations and makes the developers life simple. Here is the modified example for web intents integration for the hypothetical Open API example:

1. Register some intent upon loading our HTML document

*document.addEventListener("DOMContentLoaded", function() {*
  *var regBtn = document.getElementById("register");*

```
regBtn.addEventListener("click", function() {
  window.navigator.register("http://location/resource/subs/id", undefined);
}, false);
```

2. Start intent's activity (subscription in Open API)

```
var startButton = document.getElementById("startActivity");
startButton.addEventListener("click", function() {
  var intent = new Intent();
  intent.action = " http://location/resource/subs/id";
    window.navigator.startActivity(intent);
}, false);
```

3. Get measurements (notification in Open API) and display them in our application

```
window.navigator.onActivity = function(data) {
  var output = document.getElementById("output");
  output.textContent = JSON.stringify(data);
};
}, false);
```

Obviously, that it is much shorter than the long sequence of individual calls as per M2M Open API.

The key point here is onActivity callback that returns JSON (not XML!) formatted data. As per suggested M2M API we should perform several individual requests, parse XML responses for the each of them and only after that make some visualization. Additionally, web intents based approach is asynchronous by its nature, so, we do not need to organize asynchronous calls by our own.

Also, Web Intents approach lets us bypass sandbox restrictions. In other words, developers can raise requests right from the end-user devices, rather than always call the server. The server-side only solution becomes bottleneck very fast. And vice-versa, client side based requests let developers deploy new services very quickly. Why do not use the powerful browsers in the modern smart-phones? At the end of the day Parlay specifications were born in the time of WAP and very rudimentary mobile browsers. Why do we ignore HTML5 browsers and JavaScript support in the modern phones?

## 8. Conclusion

The paper discusses M2M communications standards for smart metering. We point out the failure of ETSI standartization process for M2M communications. We promote M-Bus protocol as a unique protocol between tens of existing ones and, therefore, support Open Metering System based on M-Bus. We point out the problem - how to estimate wireless M-bus throughput and how to avoid collisions. After analysis of Open API for M2M, submitted to ETSI, we offer a new approach in the client-side web development – Web Intents. The main goal for our suggestions is to simplify the development phase for new applications by support asynchronous calls and JSON versus XML.

## 9. Acknowledgement

The paper is financed from The Ministry of education and science of Russia (project 2011-1.9-519-0030169, contract 11.519.11.6007) being implemented in cooperation between National Research University of Electronic Technology (Zelenograd, Moscow) and Riga Technical University (Latvia)